\documentclass{article}

\usepackage{arxiv}
\usepackage{comment}

\usepackage[utf8]{inputenc} 
\usepackage[T1]{fontenc}    
\usepackage[hyphens,spaces,obeyspaces]{url}
\usepackage{booktabs}       
\usepackage{amsfonts}       
\usepackage{nicefrac}       
\usepackage{microtype}      
\usepackage{lipsum}
\usepackage{graphicx}
\usepackage{subfigure}
\usepackage{enumitem}
\usepackage{xcolor}
\graphicspath{ {./images/} }

\title{{\m}: COVID-19 Healthcare Misinformation Dataset}

\author{
 Limeng Cui\hspace{0.5in} Dongwon Lee  \vspace{0.1in} \\
  The Pennsylvania State University\\
  University Park, PA 16802
\vspace{0.1in} \\
  \texttt{\{lzc334, dongwon\}@psu.edu} \\
}

\newcommand{\lee}[1]{\textcolor{magenta}{\emph{[Dongwon: #1]}}}
\newcommand{\m}{{\sf {CoAID}}} 
\newcommand{\co}{{\sc {COVID-19}}}

\begin{document}
\maketitle
\begin{abstract}
As the {\co} virus quickly spreads around the world, unfortunately, misinformation related to {\co} also gets created and spreads like wild fire. Such misinformation has caused confusion among people, disruptions in society, and even deadly consequences in health problems.
To be able to understand, detect, and mitigate such {\co} misinformation, therefore, has not only deep intellectual values but also huge societal impacts. To help researchers  combat {\co} health misinformation, therefore, we present {\m} (\underline{Co}vid-19 he\underline{A}lthcare m\underline{I}sinformation \underline{D}ataset), with diverse {\co} healthcare misinformation, including fake news on websites and social platforms, along with users' social engagement about such news. {\m} includes 4,251 news, 296,000 related user engagements, 926 social platform posts about {\co}, and ground truth labels. The dataset is available at: \url{https://github.com/cuilimeng/CoAID}.
\end{abstract}


\maketitle

\section{Introduction}
{\co} is believed to be caused by a novel coronavirus called SARS-CoV-2, which was initially discovered in Wuhan, China in December 2019, and has quickly spread throughout the world. WHO declared the outbreak a Public Health Emergency of International Concern on January 30, 2020 \cite{world2020statement} and characterized {\co} as a pandemic on March 11, 2020 \cite{world2020directora}. As of May 6, 2020, more than 3.5 million cases of {\co} and almost 250,000 deaths worldwide have been reported to WHO \cite{world2020directorb}.
Common symptoms of {\co} includes cough, shortness of breath, fever, sore throat, and loss of taste or smell \cite{centers2020s}. The incubation period is estimated to extend to 14 days, with a median time of 5.1 days \cite{lauer2020incubation}. Mortality for {\co} (3-4\%) appears higher than that for seasonal influenza (<0.1\%) \cite{world2020qa}. Until now, there has been no known vaccine or specific cure for {\co} \cite{world2020qb}.

While {\co} situation has gotten severely worse, misinformation related to 
{\co} has rapidly risen and caused serious social disruptions as well. One the one hand, fake cures for {\co} have seriously threatened people's lives. For example, an Arizona man was dead and his wife was hospitalized after the couple ingested a form of Chloroquine to prevent {\co}\footnote{\url{https://www.npr.org/sections/coronavirus-live-updates/2020/03/24/820512107/man-dies-woman-hospitalized-after-taking-form-of-chloroquine-to-prevent-covid-19}}. On the other hand, the prevalent misinformation is disrupting social order. For example, 77 cell phone towers have been set on fire due to the conspiracy that 5G mobile networks can spread {\co}\footnote{\url{https://www.businessinsider.com/77-phone-masts-fire-coronavirus-5g-conspiracy-theory-2020-5}}.

In recent years, in mitigating these problems, many computational methods have been developed to auto-detect diverse types of misinformation in different genres
\cite{ruchansky2017csi, wang2018eann, shu2019defend, cui2019same}. However, debunking {\co}-related misinformation exhibits its own set of unique challenges. First, aided by the fear of the unknown, misinformation about emerging diseases can spread more quickly than ordinary misinformation does before being debunked\footnote{\url{https://www.jhunewsletter.com/article/2020/04/opposing-viewpoints-how-the-media-contributes-to-misinformation-in-crisis}}. Second, when one keeps seeing a piece of fake news in the news feed, she tends to think that it is true, even if she had doubts before\footnote{\url{https://www.bbc.com/future/article/20200406-why-smart-people-believe-coronavirus-myths}}. Third, after people were convinced by misinformation, ``myths correction message'' from authority may be ineffective or even reduce people's beliefs in other facts about an epidemic \cite{carey2020effects}.
Therefore, it is not clear how easy or difficult for researchers to build satisfactory computational models to auto-detect {\co}-related misinformation.

To aid these computational efforts, we have constructed a benchmark dataset, named as {\m} (\underline{Co}vid-19 he\underline{A}lthcare m\underline{I}sinformation \underline{D}ataset).
{\m} includes confirmed fake and true news articles from the fact-checked or reliable websites and posts at social platforms.
We also conducted quick data analysis to demonstrate the utility of {\m}. We contrasted the distinctive features between misinformation and facts about {\co}, and ran several state-of-the-art misinformation detection methods on {\m} to show where solutions currently lie. Our aim is to call out for public attention to the problems of {\co}-related misinformation and work together to help develop accurate detection and deterrence of such misinformation.

\section{Preliminaries}


According to the Oxford English Dictionary, both \emph{misinformation}\footnote{\url{https://www.oed.com/view/Entry/119699?redirectedFrom=misinformation}} and \emph{disinformation}\footnote{\url{https://www.oed.com/view/Entry/54579?redirectedFrom=disinformation}} are either wrong or misleading information, but disinformation is spread ``deliberately" while misinformation is not necessarily. 
Literature provides more fine-grained definitions of various examples of misinformation on social media, including rumor, fake news, hoax, satire, propaganda, and even conspiracy~\cite{shao2016hoaxy,Maria19}, such that: 
(1) \emph{fake news} is the presentation of fake or misleading claims as news, where the claims are misleading deliberately \cite{gelfert2018fake}, (2) \emph{hoax} is deliberately fabricated falsehood, with the intention to deceive a certain group of the population \cite{gelfert2018fake}, (3) \emph{rumor} is unverified but relevant information that arise in contexts such as danger or potential threat, that helps people make sense and manage risk \cite{difonzo2007rumor}. 

In this work, we do not differentiate between misinformation and disinformation as it is virtually impossible to computationally determine one's intention. We also interchangeably use fake news and misinformation in this work. {\m} comprehensively collects various misinformation examples across different platforms.

\section{Related Work}


\begin{table}[]
\centering
\caption{Comparison with existing misinformation datasets}
\label{tab:comparison}
\begin{tabular}{lccccc}
\toprule
            & \multicolumn{3}{c}{Information Type} & \multicolumn{2}{c}{User Engagement} \\ \cline{2-6} 
            & Claim        & News Article        & Post        & Tweet            & Reply            \\\midrule
LIAR        &-              &$\surd$                &-             &-                  &-                  \\
FA-KES      &-              &$\surd$                &-             &-                  &-                  \\
FakeNewsNet &-              &$\surd$                &-             &$\surd$                  &$\surd$                  \\
FakeHealth  &-              &$\surd$                &-             &$\surd$                  &$\surd$                 \\
{\m}  &$\surd$              &$\surd$                &$\surd$             &$\surd$                  &$\surd$                 \\\bottomrule
\end{tabular}
\end{table}

The construction of fake news dataset aim to extract useful features that can better distinguish fake news from true ones. Researchers have proposed a series of methods to extract news and verify the truthfulness of the news. There exists several benchmark datasets for fake news detection, that contain the linguistic features of news.

LIAR \cite{wang2017liar} is collected from political fact-checking website PolitiFact. It has 12.8K human labeled short statements from PolitiFact API, and each statement includes statement content, speaker and context. The truthfulness of each statement is evaluated by an editor from PolitiFact, by using detailed judgments and  fine-grained labels: \textit{pants-fire}, \textit{false}, \textit{barely true}, \textit{half-true}, \textit{mostly-true}, and \textit{true}.

FA-KES \cite{salem2019fa} includes the news articles around the Syrian war. The authors first identified the major events in the Syrian war from VDC (Violations Documentation Center), and then retrieved 804 news articles related to the events from the following several news outlets. The truthfulness of each article is justified by comparing the articles against the VDC dataset.

Dhoju et al. \cite{dhoju2019differences} proposed a method to collect a healthcare news dataset. They first identified reliable (WHO, CDC/NIH/NCBI, mainstream media outlets) and unreliable media outlets (Dr. Melissa Zimdars's list, fact-checking websit snopes.com), and then used news article scraping tool Newspaper3k\footnote{\url{https://newspaper.readthedocs.io/en/latest/}} to crawl the news articles. Finally they used Google Cloud Natural Language API to filter out non-health news articles.

Besides the news content, other researchers also collected user engagement features of the news, such as user engagement on online social media.

Ghenai et al. \cite{ghenai2018fake} collected 126 unproven cancer treatments that are scrutinized by an oncologist from medical sources DC Science, Wikipedia and PubMed. Topic keywords were generated from the treatments and used to find related tweets.

FakeNewsNet \cite{shu2018fakenewsnet} is collected from fact-checking websites PolitiFact and GossipCop. News contents and the related ground truth labels are crawled from the two websites. And the author collected social engagement through Twitter's Advanced Search API. The dataset includes news ID, URL, article title, tweet IDs of tweets sharing the news and scripts to crawl news contents and social engagement. It contains 1,056 articles from PolitiFact and 22,864 articles from GossipCop. Each article is labeled as fake or true.

FakeHealth \cite{dai2020ginger} is collected from healthcare information review website Health News Review, which reviews whether a news is satisfactory from 10 criteria. Daily publication of new content on this website was ceased at the end of 2018. The authors crawled users' replies, retweets and profiles by using article URL and article title by using Twitter API.

In addition, existing datasets often focus only on the news coverage or articles, and little attention has been paid to social media posts. However, fake social media posts can easily go viral on multiple social platforms, and cause serious social disruptions. For example, a fake video on YouTube claiming that a US patent on the coronavirus was applied in 2006, quickly went viral on multiple social media platforms\footnote{\url{https://factcheck.afp.com/false-claims-patents-fuel-novel-coronavirus-conspiracy-theories-online}}, causing much confusion and discomfort.

As shown in Table \ref{tab:comparison}, {\m} dataset not only includes both true and fake news, but also has short claims and social media posts. 
Compared with news articles, a \emph{claim} is much shorter--only one or two sentences--such as ``Eating garlic prevents infection with the new coronavirus.'' In this work, we term ``myth'' and ``rumor'' as ``fake claim'', in order to contrast those with ``true claim.'' Our dataset collects all possible features of interests, including multi-modal information and user engagement data. In addition, {\m} can be updated automatically to fetch the latest news. The statistics of  {\co} news dataset is shown in Table \ref{tab:dataset}.

\section{Dataset Construction}

In this section, we introduce how we have collected healthcare misinformation related to {\co},  their reliable ground truth labels, and associated user engagement features, and how we manage to update the dataset automatically.

\subsection{Facts and Misinformation on Websites}

\begin{figure}[tb]
\centering
\subfigure[A news article at the Healthline.com website]{
\label{fig:screenshot:news}
\includegraphics[height=2.5in]{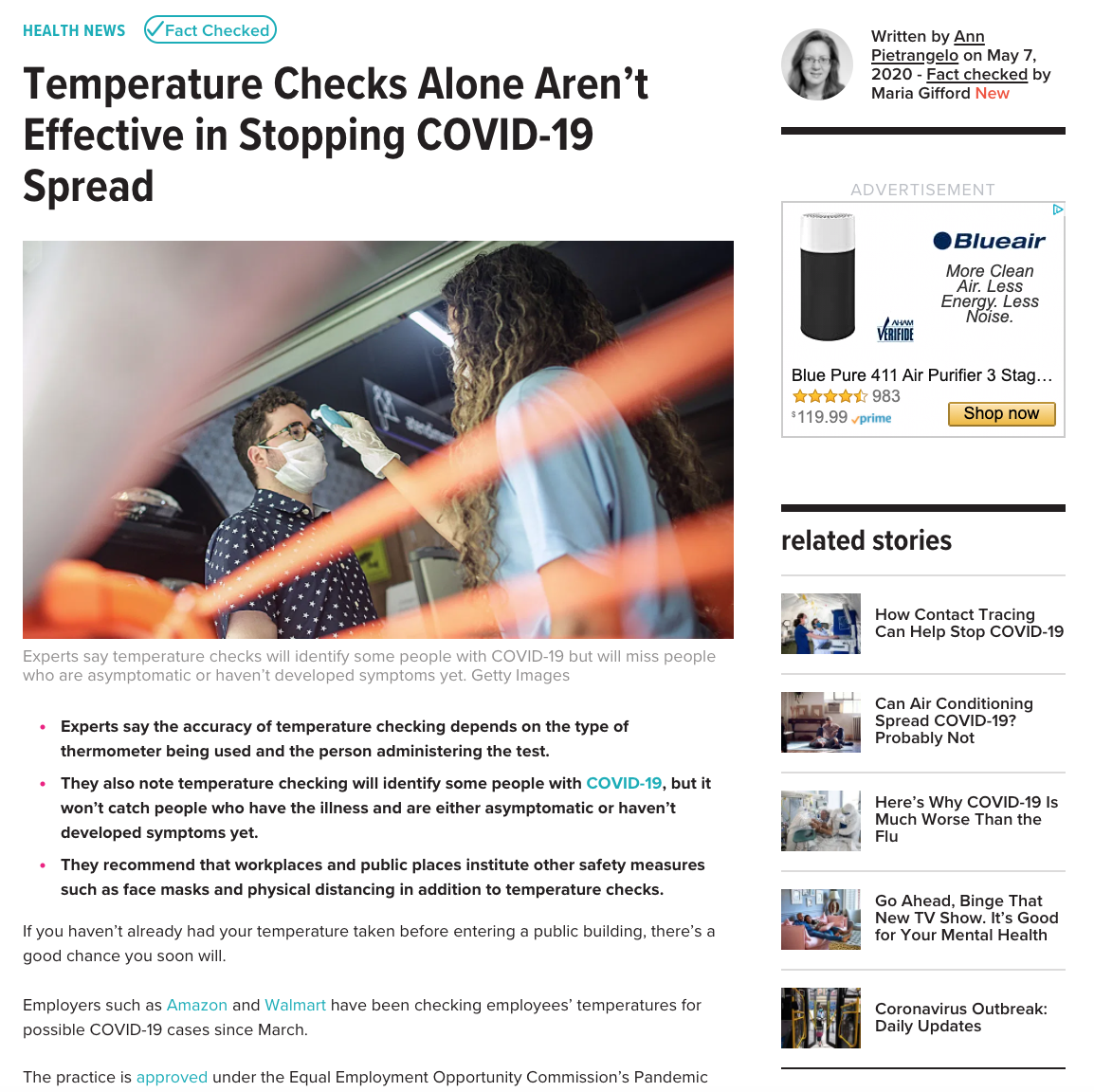}}
\hspace{0.2in}
\subfigure[Tweets]{
\label{fig:screenshot:user}
\includegraphics[height=2.5in]{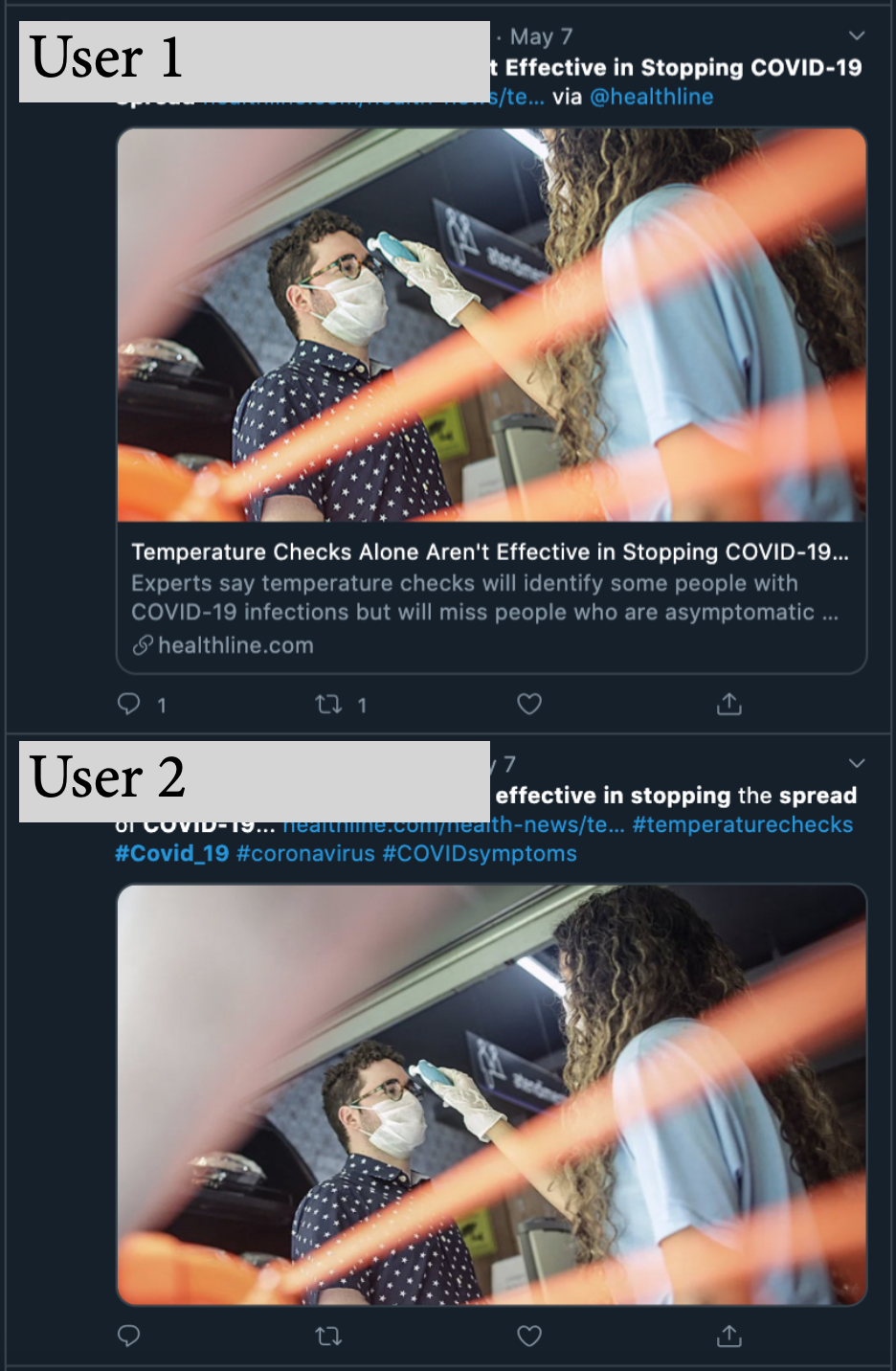}}
\hspace{0.2in}
\subfigure[Replies]{
\label{fig:screenshot:userreplies}
\includegraphics[height=2.5in]{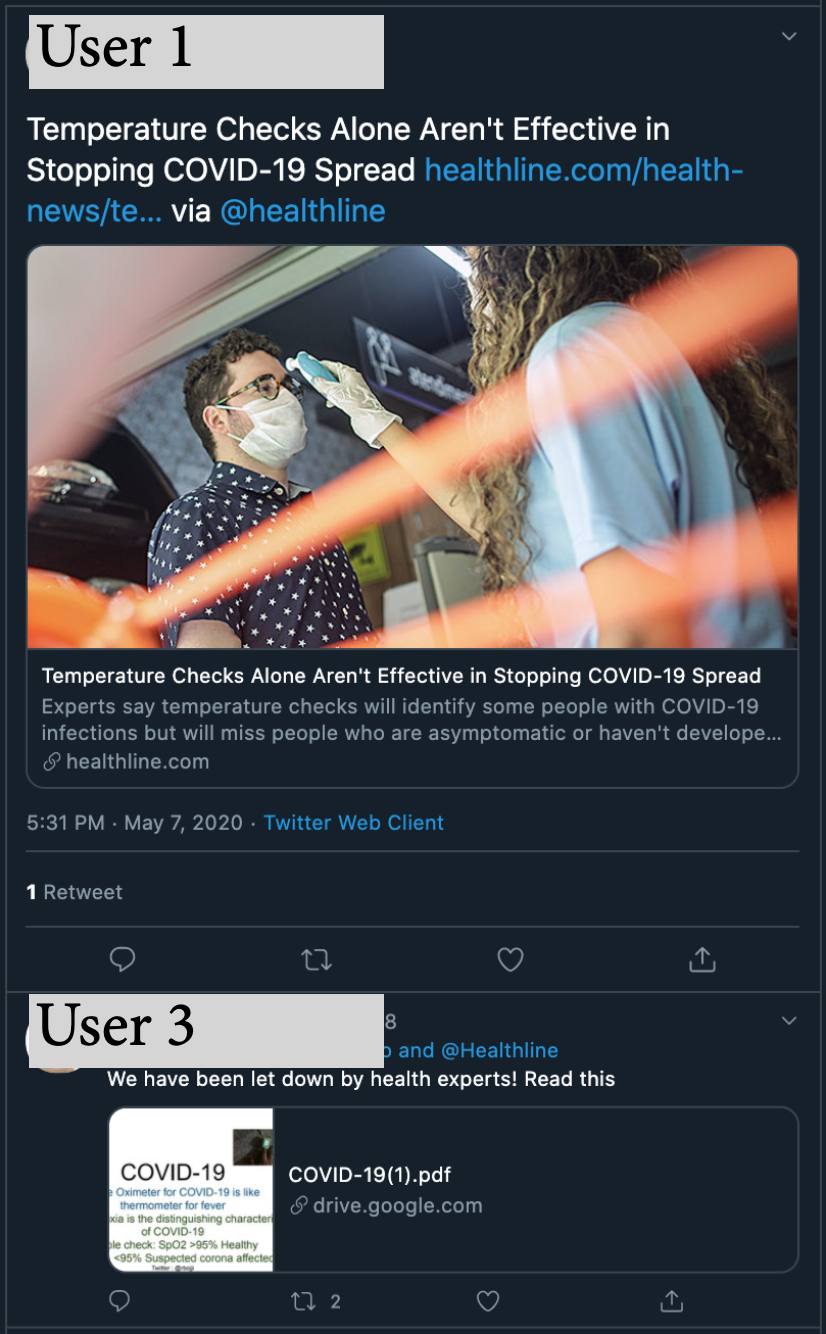}}
\caption{Screenshots of (a) a news article at the Healthline.com website and its related user engagement, including (b) tweets and (c) replies of each tweet.}
\label{fig:screenshot} 
\end{figure}

\begin{figure}[tb]
  \centering
  \includegraphics[width=0.8\linewidth]{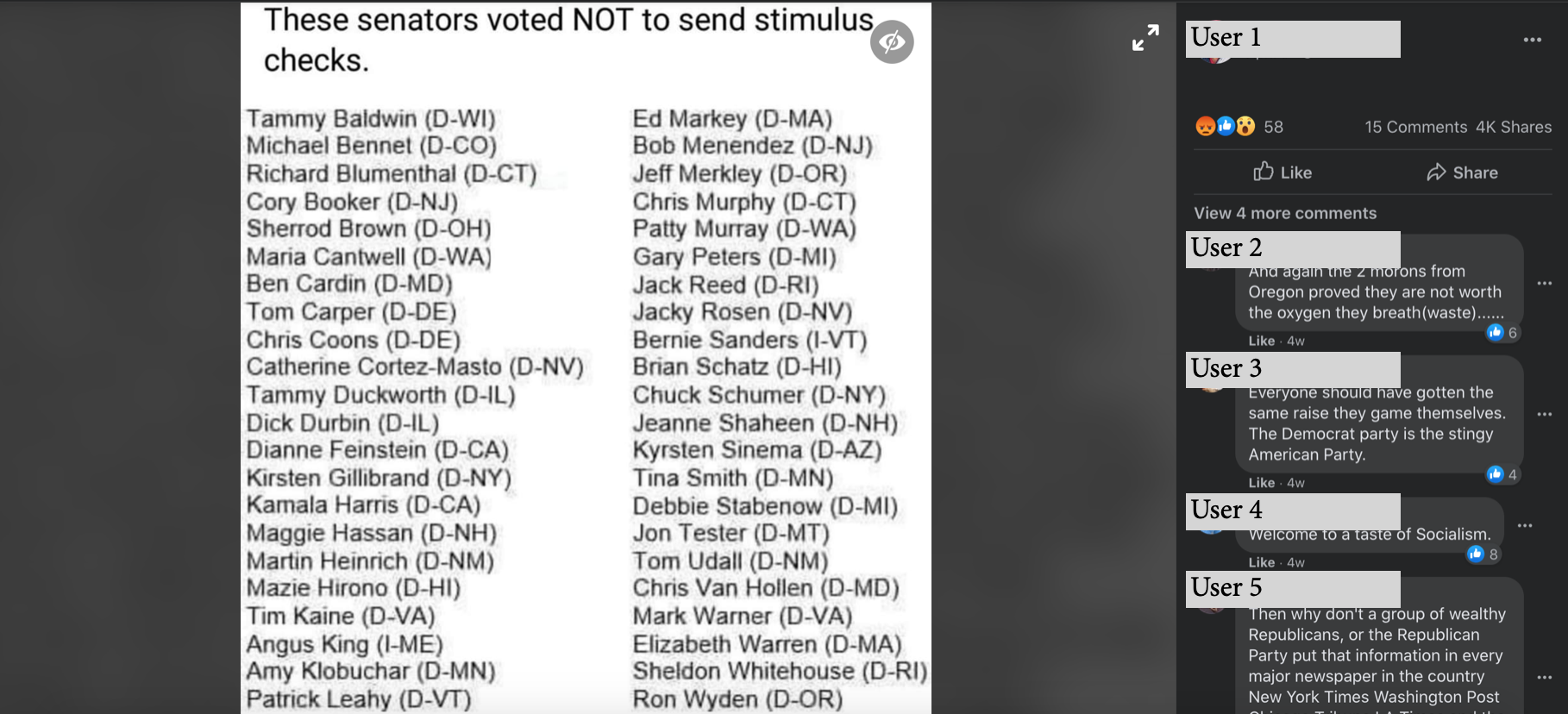}
  \caption{Screenshot of a social platform post.}
\label{fig:socialpost}
\end{figure}

We have collected both facts and misinformation related to {\co}, including news articles and claims. The publishing dates of the collected information range from December 1, 2019 to September 1, 2020. The topics include: \textit{{\co}}, \textit{coronavirus}, \textit{pneumonia}, \textit{flu}\footnote{We filtered out known flu types such as \textit{Influenza A/B}, \textit{bird flu} and \textit{swine flu}.}, \textit{lock down}, \textit{stay home}, \textit{quarantine} and \textit{ventilator}.

\begin{itemize}

\item \textbf{News Article}: An example of news article is shown in Figure \ref{fig:screenshot:news}. We identified reliable media outlets and fact-checking websites and obtained URLs of misinformation (i.e., fake news) and facts (i.e., true news). To collect true news articles, 
we have crawled from 9 reliable media outlets that have been cross-checked as reliable, including: Healthline\footnote{\url{https://www.healthline.com/health-news?ref=global}}, ScienceDaily\footnote{\url{https://www.sciencedaily.com/news/health_medicine/}}, NIH\footnote{\url{https://www.nih.gov/news-events/news-releases}} (National Institutes of Health), MNT\footnote{\url{https://www.medicalnewstoday.com/coronavirus}} (Medical News Today), Mayo Clinic\footnote{\url{https://www.mayoclinic.org/diseases-conditions/coronavirus/symptoms-causes/syc-20479963}}, Cleveland Clinic\footnote{\url{https://newsroom.clevelandclinic.org/}}, WebMD\footnote{\url{https://www.webmd.com/news/articles}}, WHO\footnote{\url{https://www.who.int/news-room}} (World Health Organization) and CDC\footnote{\url{https://www.cdc.gov/coronavirus/2019-ncov/whats-new-all.html}} (Centers for Disease Control and Prevention). 
We used the Selenium\footnote{\url{https://www.selenium.dev/}} to parse the html structure of web pages to locate the URLs. 
To collect fake news, we retrieved the URLs from several fact-checking websites, including: LeadStories\footnote{\url{https://leadstories.com/hoax-alert/}}, PolitiFact\footnote{\url{https://www.politifact.com/coronavirus/}}, FactCheck.org\footnote{\url{https://www.factcheck.org/fake-news/}}, CheckYourFact\footnote{\url{https://checkyourfact.com/}}, AFP Fact Check\footnote{\url{https://factcheck.afp.com/}} and Health Feedback\footnote{\url{https://healthfeedback.org/}}.

\item \textbf{Claim}: To collect claims (of one or two sentences), 
we referred to the  WHO official website\footnote{\url{https://www.who.int/}}, WHO official Twitter account\footnote{\url{https://twitter.com/WHO}} and MNT\footnote{\url{https://www.medicalnewstoday.com/articles/coronavirus-myths-explored}}. We specifically separated true and fake claims. For example, ``Only older adults and young people are at risk'' is a fake claim, while ``5G mobile networks DO NOT spread {\co}'' is a true claim.
\end{itemize}

After we obtained all URLs to true and fake news related to {\co}, 
we used the Newspaper3k\footnote{\url{https://newspaper.readthedocs.io/en/latest/}} to fetch their corresponding title, content, abstract, and keywords. In total, we have obtained 204 fake news articles, 3,565 true news articles, 28 fake claims and 454 true claims. 

\subsection{User Engagement}

\begin{table*}[tb]
\centering
\caption{Statistics of {\m} Version 0.3}
\label{tab:dataset}
\begin{tabular}{lcccc}
\toprule
                         & \multicolumn{2}{c}{Fake}  & \multicolumn{2}{c}{True}   \\ \cline{2-5}
                         & Claim    & News Article & Claim      & News Article \\\midrule
\# Information on Website       & 28            & 204           & 454             & 3,565         \\
\# Tweets                & 484           & 10,439         & 8,092           & 141,652        \\
\# Replies               & 626           & 7,436         & 12,451           & 114,820        \\\midrule
\# Social Platform Posts & 650 (880)     & -             & 42 (46)         & -             \\\bottomrule
\# Total                 & 1,788 (2,018) & 18,079        & 21,039 (21,043) & 260,037      \\\bottomrule
\end{tabular}
\end{table*}

We used Twitter API to fetch user engagement data of both facts and misinformation. Figure \ref{fig:screenshot:user} shows user tweets related to the news in Figure \ref{fig:screenshot:news}, and Figure \ref{fig:screenshot:userreplies} shows user replies under a tweet.

\begin{itemize}
     
\item \textbf{Tweets}: We used the titles of news articles as the search queries, and specified start and end dates to get the tweets discussing about the news in question in a certain period. The retrieved user engagement features include: user ID, tweets, replies, favorites, retweets, and location. In total, we obtained 10,439 tweets about fake articles, 141,652 tweets about true news articles, 484 tweets about fake claims and 8,092 tweets about true claims, which is summarized in Table \ref{tab:dataset}.

\item \textbf{Replies}: We further obtained users' replies of each tweet, by using  tweet IDs. In total, we obtained 7,436 replies under fake articles, 114,820 replies under true news articles, 626 tweets about fake claims and 12,451 tweets under true claims, which is summarized in Table \ref{tab:dataset}.
\end{itemize}

\subsection{Social Platform Posts}
In addition to traditional mass media, on social media, individual users act as the producers of user generated contents, forming the so-called self media.
An example of such self-media is shown in Figure \ref{fig:socialpost}. As both true and fake news spread through self-media as well, we have also collected both fake and true posts originated from five well-known social media platforms--Facebook, Twitter, Instagram, Youtube, and TikTok--also fact-checked by: Lead Stories, PolitiFact, FactCheck.org, CheckYourFact, AFP Fact Check, and Health Feedback. We only count once for each distinctive fake post. For example, if a post was originally posted by a Facebook user and then posted again by another user on  Twitter, it is counted as 1 (i.e., no duplicate fake posts). The numbers in parentheses in Table \ref{tab:dataset} represent the total number of posts, before duplicates were removed.

Per each post, we only crawled its title without news content, and listed the posts under ``Claim'' in Table \ref{tab:dataset}.

\subsection{Automatic Updates}

{\m} can be updated automatically with the latest news and tweets. We record the timestamp of the most recently added data each time, and use it as the start date for the next search.

\vspace{0.1in}
In summary, we list the description of the  extracted features  in Table \ref{tab:features}. Note that
``Title'' is the title of the related fact-checking article, while ``Article title'' is the title of the article crawled. 
Due to the limitation under Twitter's Terms \& Conditions\footnote{\url{https://developer.twitter.com/en/developer-terms/agreement-and-policy}}, only tweet IDs are included in the released dataset of {\m}.

\begin{table}[tb]
\centering
  \caption{Feature descriptions of {\m}}
  \label{tab:features}
  \begin{tabular}{p{0.25\linewidth}p{0.65\linewidth}}
    \toprule
    Type & Features \\
    \midrule
    Information on Websites & ID, Fact-checking URL, Information URLs, Title, Article title, Content, Abstract, Publish date, Keywords\\
    User engagement: Tweets & ID, Tweet ID\\
    User engagement: Replies & ID, Tweet ID, Reply ID\\
    Social Platform Posts & ID, Fact-checking URL, Post URLs, Title \\
  \bottomrule
\end{tabular}
\end{table}

\section{Data Analysis}

{\m} includes multi-modal news, related ground truth labels, and user engagement. The detailed statistics of the dataset are shown in Table \ref{tab:dataset}. In this section, we perform preliminary data analysis in order to illustrate the features of {\m}. Then, we perform several fake news detection methods to evaluate the challenges of {\co} healthcare misinformation detection. The following data analysis is based on {\m} Version 0.1 in Table \ref{tab:datasetv1}.

\subsection{Prescriptive Analysis}



\begin{figure}[tb]
\centering
\subfigure[Fake News]{
\label{fig:sentiment:fake}
\includegraphics[width=0.48\columnwidth]{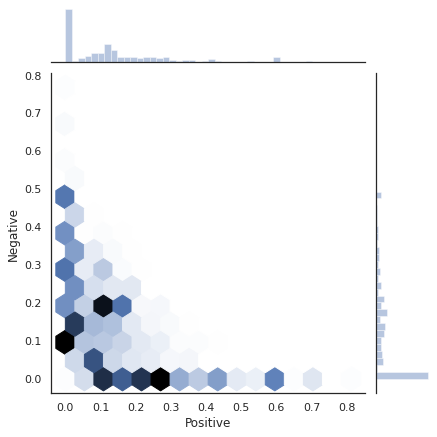}}
\subfigure[True News]{
\label{fig:sentiment:real}
\includegraphics[width=0.48\columnwidth]{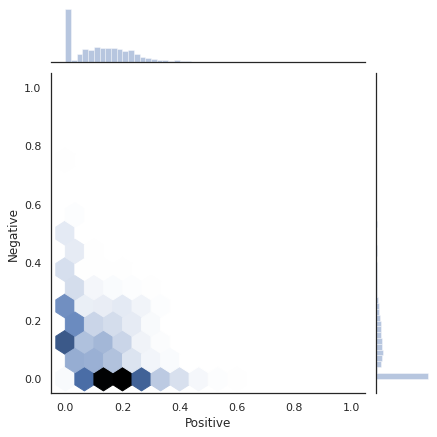}}
\caption{Sentiments of tweets related to fake and true news articles} 
\label{fig:sentiment} 
\end{figure}

We use VADER \cite{hutto2014vader} to analyze users' sentiments in tweets related to fake and true news articles. 
We exclude the tweets that are completely neutral, and plot the sentiment scores of tweets related to fake articles in Figure \ref{fig:sentiment:fake} and true news articles in Figure \ref{fig:sentiment:real}, respectively. The bars show the numbers of tweets with different negative and positive degrees, while the scatter plot shows the negative and positive scores. The intensity of the hive represents the density of data points. We can see from Figure \ref{fig:sentiment} that tweets related to fake news are more negative, and have stronger sentiment polarities than those related to true news articles.

\begin{figure}[tb]
\centering
\subfigure[Fake News]{
\label{fig:hashtag:fake}
\includegraphics[width=0.48\columnwidth]{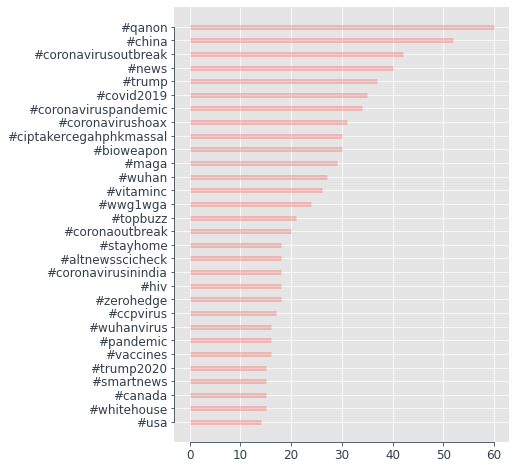}}
\subfigure[True News]{
\label{fig:hashtag:real}
\includegraphics[width=0.48\columnwidth]{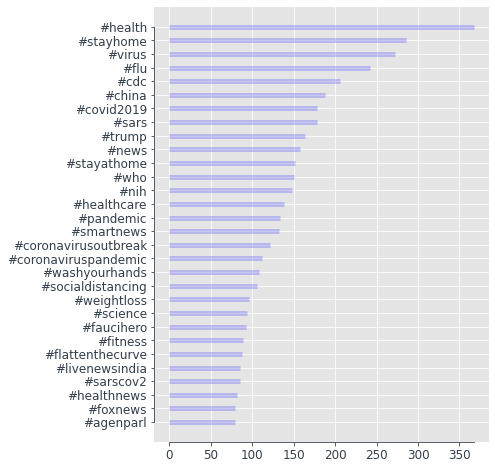}}
\caption{Frequency of hashtags in tweets about fake and true news articles} 
\label{fig:hashtag} 
\end{figure}

Next, we analyze the top 30 frequent hashtags in tweets related to fake and true news articles, respectively. For a more intuitive view, we delete the most frequent hashtags ``\#coronavirus'', ``\#covid\_19'', ``\#covid19'' and ``\#covid'' as they appear almost in every tweet. We show the frequency of hashtags in tweets related to fake and true news articles in Figure \ref{fig:hashtag:fake} and Figure \ref{fig:hashtag:real}, respectively. We find that 
the hashtag distributions of tweets about fake and true news articles are quite different. While the hashtags in tweets about true news articles are mainly related to healthcare, those in tweets about fake news cover more diverse topics, including conspiracy (\#bioweapon) and fake cure (\#vitaminc).

\begin{figure}[tb]
  \centering
  \includegraphics[width=0.9\linewidth]{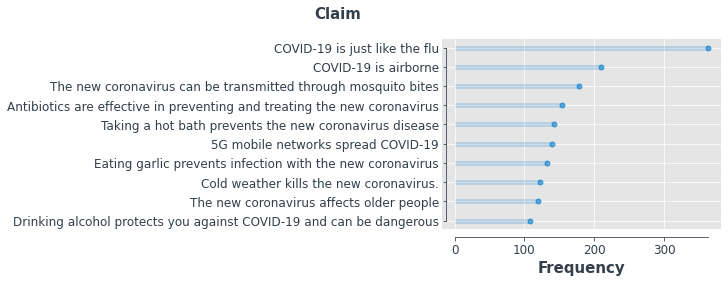}
  \caption{Frequency of common claims}
\label{fig:questions}
\end{figure}



\begin{figure}[tb]
\centering
\begin{subfigure}[Daily counts of tweets related to ``{\co} is just like the flu'']{
\includegraphics[width=0.9\linewidth]{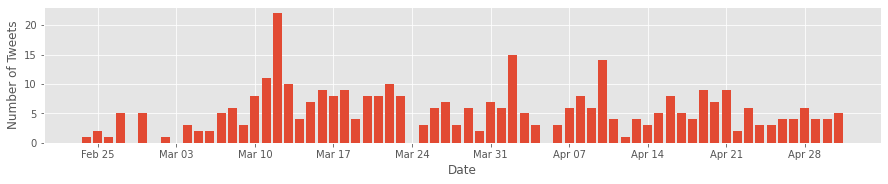}   
\label{fig:dailycounts:justaflu} }
\end{subfigure}
\hfill
\begin{subfigure}[Daily counts of tweets related to ``5G mobile networks spread {\co}'']{
\includegraphics[width=0.9\linewidth]{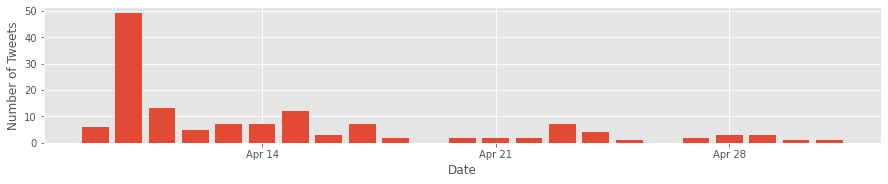}
\label{fig:dailycounts:5g} }%
\end{subfigure}
\label{myfigure}                            
\caption{Claim trends}
\end{figure}

In Figure \ref{fig:questions}, next, we list ten most common claims of {\co} on Twitter, and their corresponding frequencies. We merge similar claims into one, such as ``Eating garlic prevents infection with the new coronavirus'' and ``Garlic protects against coronaviruses.'' We can see that the most frequently discussed claim is ``{\co} is just like the flu'', followed by ``{\co} is airborne'' and ``The new coronavirus can be transmitted through mosquito bites''.
We then plot the daily counts of tweets related to two claims, `{\co} is just like the flu'' and ``5G mobile networks spread {\co}'' in Figure \ref{fig:dailycounts:justaflu} and \ref{fig:dailycounts:5g}, respectively. We can see that there were most tweets about ``{\co} is just like the flu'' at the beginning of the global outbreak, which is around March 12, 2020. Note that this topic is mentioned almost everyday. The tweets about ``5G mobile networks spread {\co}'' were mostly on the April 10, 2020, when 5G towers were being set on fire, but afterward the number declined as the claim was debunked.

\subsection{{\co} Misinformation Detection}
To demonstrate the main utility of {m}, here, we conduct comparative experiments on the misinformation detection task. We include both simple methods and state-of-the-art methods as baselines.
We consider the following baseline methods for {\co} misinformation detection:
\begin{itemize}[leftmargin=*,noitemsep]
    \item \textbf{SVM (Support Vector Machine)}: We use bag-of-words to represent texts, and feed the representations to a linear kernel SVM.
    \item \textbf{LR (Logistic Regression)}: We concatenate all the word embeddings together, and feed it to the model.
    \item \textbf{RF (Random Forest)}: We use bag-of-words to represent texts, and feed the representations to the model.
    \item \textbf{CNN}: We feed the word embeddings into a Convolution1D, which will learn filters. Then we add a max pooling layer.
    \item \textbf{BiGRU} \cite{bahdanau2015neural}: BiGRU gets annotations of words by summarizing information from both directions for words.
    \item \textbf{CSI} \cite{ruchansky2017csi}: CSI is a hybrid deep learning-based misinformation detection model that utilizes information from article content and user response. The article representation is modeled via an LSTM model with the article embedding via Doc2Vec \cite{le2014distributed} and user response.
    \item \textbf{SAME\textbackslash{}v} \cite{cui2019same}: SAME uses news image, content, metadata and users' sentiments for fake news detection. As the majority of news does not have a cover image, the visual part of the model is ignored, and the baseline is termed as SAME\textbackslash{}v.
    \item \textbf{HAN} \cite{yang2016hierarchical}: HAN has two levels of attention mechanisms applied at the word and sentence-level, enabling it to attend differently to more and less important content for document classification.
    \item \textbf{dEFEND} \cite{shu2019defend}: dEFEND utilizes the hierarchical attention network as HAN on article content, and a co-attention mechanism between article content and user comment for misinformation detection.
\end{itemize}

To evaluate the performance of misinformation detection algorithms, we use the following metrics, which are commonly used to evaluate classifiers in related areas: PR-AUC, Precision, Recall, and F1 score.

We implement all models with Keras. We randomly use the labels of 75\% news articles for training and predict the remaining 25\%. We set the hidden dimension of all models to $100$. The word embeddings are initialized by GloVe \cite{pennington2014glove} and the dimension of pre-trained word embeddings is set to 100.
For deep learning methods, we use Adam with a minibatch of 50 articles on the dataset, and the training epoch is set as 10. For a fair comparison, we use the cross-entropy loss for all methods. Finally, we run each method five times and report the average score in Table \ref{tab:detection}.

\begin{table}[tb]
\centering
  \caption{Misinformation detection performance
  }
  \label{tab:detection}
  \begin{tabular}{lcccc}
    \toprule
     & PR AUC & Precision & Recall & F1 \\
    \midrule
    SVM & 0.3365 & 0.4036 & 0.1322 & 0.1986 \\
    LR & 0.2871 & 0.4287 & 0.0690 & 0.1143 \\
    RF & 0.3937 & 0.6056 & 0.0581 & 0.1045 \\
    CNN & 0.8126 & 0.9653 & 0.1238 & 0.1983 \\
    BiGRU & 0.2241 & 0.7476 & 0.0524 & 0.0930 \\
    CSI & 0.3576 & 0.6814 & 0.2109 & 0.2283 \\
    SAME\textbackslash{}v & 0.7901 & 0.8922 & 0.2991 & 0.3400 \\
    HAN & 0.6824 & 0.6965 & 0.4659 & 0.5471 \\
    dEFEND & 0.7229 & 0.8965 & 0.4847 & 0.5814 \\
  \bottomrule
\end{tabular}
\end{table}

From Table \ref{tab:detection}, we can see that state-of-the-art methods perform better than simple methods, as they incorporate signals from user engagement, better capturing contextual information. However, as the dataset is quite imbalanced, the models tend to generate many fake positive cases. Thus, the recall and F1 values are far from being satisfactory. 
As the proportion of true and fake information is likely to be even more skewer, practical detection solutions need to handle this type of imbalance setting more.

In addition, with regard to healthcare news, it is often more scarce and rare for lay persons to give discriminating comments due to lack of professional knowledge. For example, WHO tweeted ``There is no evidence that regularly rinsing the nose with saline solution has protected people from infection with the new coronavirus.'' The majority of comments for the tweet are unrelated, useless, and even includes hate speech and other misinformation about {\co}. Therefore, the state-of-the-art methods  \cite{ruchansky2017csi, cui2019same, shu2019defend} that can exploit the signals from user engagement have not achieved satisfactory results.

\section{Conclusion}

In this paper, we present a comprehensive {\co} misinformation dataset {\m}, which contains news articles, user engagement, and social platform posts. We describe how we collected the dataset. In addition, we conduct brief data analysis to show the distinctive features between misinformation and facts, and demonstrate the future research directions through a fake news detection task over several state-of-the-art methods.
We hope researchers to find {\m} useful for their research and together contribute to flatten the curve of {\co}.


\section*{Acknowledgments}
We thank all healthcare providers who risk their lives for others, people staying home for others, and the kindness that connects us all.

This work was in part supported by NSF awards \#1742702, \#1820609, \#1915801, and \#1934782.

\bibliographystyle{plain}
\bibliography{sample-base}

\newpage
\appendix

\section{Releases}

\subsection{Version 0.1}
This initial dataset corresponding to the paper includes {\co} information collected from December 1, 2019 through May 1, 2020. It includes 1,896 news, 183,564 related user engagements, 516 social platform posts about {\co}, and ground truth labels.

\begin{table*}[!htb]
\centering
\caption{Statistics of {\m} Version 0.1}
\label{tab:datasetv1}
\begin{tabular}{lcccc}
\toprule
                         & \multicolumn{2}{c}{Fake}  & \multicolumn{2}{c}{True}   \\ \cline{2-5}
                         & Claim    & News Article & Claim      & News Article \\\midrule
\# Information on Website       & 27            & 135           & 166             & 1,568         \\
\# Tweets                & 457           & 9,218         & 6,342           & 87,324        \\
\# Replies               & 623           & 5,721         & 9,764           & 64,115        \\\midrule
\# Social Platform Posts & 414 (492)     & -             & 21 (24)         & -             \\\bottomrule
\# Total                 & 1,521 (1,599) & 15,074        & 16,293 (16,296) & 153,007      \\\bottomrule
\end{tabular}
\end{table*}

\subsection{Version 0.2}
Besides the original data in Version 0.1, the additional data includes {\co} information collected from May 1, 2020 through July 1, 2020. It includes 1,339 news, 111,128 related user engagements, 335 social platform posts about {\co}, and ground truth labels.

The statistics of the additional data in {\m} Version 0.2 is shown in Table \ref{tab:datasetv2}.

\begin{table*}[!htb]
\centering
\caption{Statistics of the additional data in {\m} Version 0.2}
\label{tab:datasetv2}
\begin{tabular}{lcccc}
\toprule
                         & \multicolumn{2}{c}{Fake}  & \multicolumn{2}{c}{True}   \\ \cline{2-5}
                         & Claim    & News Article & Claim      & News Article \\\midrule
\# Information on Website       & 1            & 55           & 172             & 1,111         \\
\# Tweets                & 27           & 1,198         & 1,453           & 54,224        \\
\# Replies               & 3           & 1,672         & 1,909           & 50,642        \\\midrule
\# Social Platform Posts & 197 (320)     & -             & 14 (15)         & -             \\\bottomrule
\# Total                 & 228 (351) & 2,925        & 3,548 (3,549) & 105,977      \\\bottomrule
\end{tabular}
\end{table*}

\subsection{Version 0.3}
Besides the original data in Version 0.2, the additional data includes {\co} information collected from July 1, 2020 through September 1, 2020. It includes 1,016 news, 1,308 related user engagements, 75 social platform posts about {\co}, and ground truth labels.

The statistics of the additional data in {\m} Version 0.3 is shown in Table \ref{tab:datasetv3}.

\begin{table*}[!htb]
\centering
\caption{Statistics of the additional data in {\m} Version 0.3}
\label{tab:datasetv3}
\begin{tabular}{lcccc}
\toprule
                         & \multicolumn{2}{c}{Fake}  & \multicolumn{2}{c}{True}   \\ \cline{2-5}
                         & Claim    & News Article & Claim      & News Article \\\midrule
\# Information on Website       & 0            & 14           & 116             & 886         \\
\# Tweets                & 0           & 23         & 297           & 104        \\
\# Replies               & 0           & 43         & 778           & 63        \\\midrule
\# Social Platform Posts & 39 (68)     & -             & 7         & -             \\\bottomrule
\# Total                 & 39 (68) & 80        & 1,198 & 1,053      \\\bottomrule
\end{tabular}
\end{table*}

\end{document}